\documentclass{aa}  
\usepackage{graphicx}
\usepackage{txfonts}
\usepackage{natbib}
\usepackage{balance}
\usepackage{rotating}
\usepackage{lscape}
\usepackage[a4paper]{hyperref}
\newcommand{\zabs}{$z_{\rm abs}\,$}

\newcommand{\kms}{km~s$^{-1}\,$}
\newcommand{\ms}{m~s$^{-1}\,$}

\newcommand{\daa}{$\Delta\alpha/\alpha\,$}
\newcommand{\drv}{$\Delta RV\, $}
\begin{document}
\title{UVES radial velocity accuracy from asteroid observations}
\subtitle{I. Implications for the fine structure constant variability\thanks{Based on 
observations performed at the VLT Kueyen telescope (ESO, Paranal, Chile). }  }

\author{P. Molaro\inst{1},
     S. A. Levshakov\inst{2},
        S. Monai\inst{1},
        M. Centuri\'on\inst{1},
        P. Bonifacio\inst{1,3},
        S. D'Odorico\inst{4},
         \and
        L. Monaco\inst{5}
          }
\offprints{P. Molaro}
\institute{INAF-Osservatorio Astronomico di Trieste. Via G.B. Tiepolo 11 I-34143, Trieste, Italy\\
\and 
Department of Theoretical Astrophysics, Ioffe Physico-Technical Institute,
Polytekhnicheskaya Str. 26, 194021 St. Petersburg, Russian Federation\\
\and 
CIFIST, Marie Curie Excellence Team and GEPI, Observatoire de Paris, CNRS, 
Universit\'e Paris Diderot; Place Jules Janssen 92190, Meudon, France;
Observatoire de Paris 61, CNRS  avenue de l'Observatoire, 75014 Paris, France\\
\and
European Southern Observatory, Karl-Schwarzschild-Strasse 2, D-85748 Garching bei M\"unchen, Germany\\ 
\and 
European Southern Observatory, Alonso de C—rdova 3107, Casilla 19001, Vitacura, Santiago, Chile
}

\date{Received October  .., 2007; accepted ...}

 
\abstract
{High resolution observations of the asteroids Iris and Juno have been performed 
by means of  the UVES spectrograph at the ESO VLT to obtain  the  effective 
accuracy of the  spectrograph's radial velocity. The knowledge of this quantity  
has important bearings on   studies searching for a  variability of the fine structure  
constant carried on with this instrument.}
{Asteroids provide  a  precise radial velocity reference at the level of 
1 \ms\ which allows  instrumental calibration and the recognition of small instrumental 
drifts and calibration systematics. In particular, radial velocity drifts due to non 
uniform  slit illumination and slit optical  misalignment  in the two UVES spectrograph 
arms can be investigated.}
{The position of the solar spectrum reflected by the asteroids  are  compared    
with the solar wavelength positions or with that of asteroid observations at other 
epochs or with  the  twilight    to asses  UVES instrumental accuracy .}
{Radial velocities offsets in the range  $\approx$10--50 \ms\ are generally observed  
likely due to a non uniform slit illumination.  However, no  radial velocity  patterns   
with  wavelength are detected and  the two UVES arms provide consistent radial velocities.     
These results suggest that the detected \daa variability  by Levshakov et al. (2007) 
deduced from a drift of  $-180 \pm 85$ \ms\  at \zabs =1.84, 
between two sets of  $\ion{Fe}{ii}$ lines falling in the two UVES arms  may be real 
or induced by other kinds of systematics than those investigated here. 
The proposed technique allows real time quality check of the spectrograph and should 
be followed for very accurate measurements.}
{} 
\authorrunning{P. Molaro et al.}
\titlerunning{UVES radial velocity accuracy from asteroid observations}
 \maketitle
%
\section{Introduction}

Radial velocity precision is required in several fields of astronomical research 
ranging  from the detection of exoplanets to the study of the variability of the 
fundamental physical  constants. To reveal  the presence of an orbiting planet  
dedicated spectrographs have been manufactured to achieve the  best accuracy  
in the radial velocity. With HARPS at the 3.6 m telescope a relative  precision of 1 \ms\  
or higher has been achieved when the full optical  stellar spectrum of a solar 
type star is recorded and compared in different epochs.   
Search for a possible variability of the fine structure constant, 
\daa = $(\alpha_z - \alpha)/\alpha$, at a redshift $z$,  
is  currently carried out by  measuring line shifts between different lines of absorbers   
observed in spectra of distant quasars  which show different sensitivities to $\alpha$  
(Webb et al. 1999,  Dzuba et al. 2002). QSOs are rather faint and require   
large telescopes such as VLT or  Keck combined with the  high resolution spectrographs,  
UVES and HIRES respectively. Murphy et al. (2004) claim that  \daa = $-5.7\pm1.1$ ppm 
(ppm stands for parts per million, $10^{-6}$)  by averaging over
143 absorption systems detected in HIRES/Keck telescope spectra of QSOs
in a  redshift range $0.2 < z < 4.2$  implying  that  in the past the fine 
structure constant was smaller.  On the other hand no variability has been measured 
by a different group at the VLT with UVES  adopting similar techniques 
(Quast et al. 2004, Chand et al. 2004, Levshakov et al. 2005, 2006; 
but see also Murphy et al. 2006 and Srianand et al. 2007). 
More recently  Levshakov et al. (2007) measured  a radial velocity  difference of  
$-180\pm85$ \ms\  between $\ion{Fe}{ii}$ transitions falling in the two different arms 
of the UVES  providing evidence for a variation in the fine structure constant 
\daa = $5.4\pm2.5$ ppm, with the fine structure constant  being larger in the past 
at odds with what found by  Murphy et al. (2004).  
Given the importance of these results for fundamental physics a thorough 
investigation of systematic errors  to rule out possible instrumental shifts which 
may occur during UVES observations  is rather crucial.  

Spectroscopic observations are  generally calibrated in wavelength by means  of standard 
calibration lamps, namely the ThAr lamps. However,  to achieve a  \daa\  of 1~ppm   
a precision of 30 \ms\ in the radial velocity of the most sensitive lines is required 
challenging the spectrograph precision. Small instrumental effects could  be present since 
the light paths of calibration and stellar beams are different when entering the spectrograph slits. 
Instrumental flexures, temperatures and atmospheric pressure instability can produce small radial
velocity shifts between  calibration and  science observations.
Temperature and pressure variations   as small as  
$\Delta T = 0.3$~K or a $\Delta P = 1$~mbar produce a drift  of $\approx$50 \ms\  
(Kaufer et al. 2004). These effects  can be
minimized  by taking ThAr lamps immediately before or after  the science
exposures if  ambient conditions do not change in the meantime.  
However,  an  uneven illumination of the slit  may cause spectral shifts and 
therefore errors in the measurements of radial velocities. This problem is particularly 
acute in the case of UVES observations with the dichroic mode   
where  the light enters two distinct slits of the two arm spectrograph. 
Possible  effects of different  illumination of the two slits of the blue and red arms of UVES
are unknown.  

To probe small possible instrumental effects in UVES,  we observed the solar spectrum 
reflected by  asteroids which are sources with radial velocities known at the \ms\ level. 
This accuracy is not required by the majority of the observations but is  crucial  for
the investigation of variability of the fine structure constant. The presence of 
a variability of fundamental dimensionless constants would be a discovery of the 
outmost importance in theoretical physics with far reaching implications 
(Copeland et al. 2006, Avelino et al. 2006, Martins 2006, Fujii  2007).

\begin{table*}[]
\caption{\footnotesize{Journal of asteroid  observations and basic data. 
Ceres spectra has been observed by HARPS. Expected radial velocities and 
its components are given in columns 6, 7 and 8. For Ceres the values refer to mid exposure.}} 
\label{obsdata}
\begin{tabular}{lccccccc}
\hline
\hline
Name & Date     &  JD            &  $V$   & exp.& $RV_{ast-par}$ & $RV_{ast-\sun}$ &$\Delta RV$ \\ 
     &          &                &  mag     & sec &    \kms        &     \kms        &   \kms    \\
\hline
Iris& 18/12/06 & 2454088.518086 &  7.83 & 300 & 12.707 & 1.704   & 14.411 \\
     & 22/12/06 & 2454092.514539 &  7.95 & 300 & 13.777 & 1.830   & 15.607 \\
     & 23/12/06 & 2454093.514000 &  7.98 & 300 & 14.030 & 1.862   & 15.892 \\
     & 24/12/06 & 2454094.515400 &  8.01 & 300 & 14.283 & 1.893   & 16.176 \\
     & 25/12/06 & 2454095.513566 &  8.04 & 450 & 14.521 & 1.924   & 16.445 \\
Juno & 24/01/07 & 2454125.880451 & 10.62 & 600 & -21.403 & 3.731  & -17.672 \\
     & 25/01/07 & 2454126.873002 & 10.61 & 900 & -21.364 & 3.721  & -17.643 \\
     & 29/01/07 & 2454130.873891 & 10.57 & 900 & -21.065 & 3.683  & -17.382 \\
     & 31/01/07 & 2454132.846757 & 10.55 & 900 & -20.949 & 3.664  & -17.285 \\
Ceres& 15/07/06 & 2453932.837256 &  8.00 & 1800& -11.288 & 0.456  & -10.832 \\
     & 22/05/06 & 2453877.919808 &  8.85 & 900 & -22.707 & 0.690  & -22.017 \\
\hline
\end{tabular} 
\end{table*}

\begin{table*}[]
\caption{\footnotesize{Sky observations and data, *spectra taken with  HARPS}} 
\label{obsdatas}
\begin{tabular}{lcc}
\hline
\hline
 Date     &  JD     & $RV_{par-\sun}$ \\ 
          &         &  \kms    \\
\hline
 18/12/06 & 2454088.480700 & 0.238 \\
 22/12/06 & 2454092.481400 & 0.263 \\
 23/12/06 & 2454093.479390 & 0.272 \\
 24/12/06 & 2454094.478500 & 0.279 \\
 25/12/06 & 2454095.480300 & 0.286 \\
 25/01/07 & 2454126.472293 & 0.573 \\
 31/01/07 & 2454132.482003 & 0.629 \\
 14/07/06* & 2453931.425152 & 0.302 \\
 22/10/05* & 2453666.426383 & -0.070 \\
\hline
\end{tabular} 
\end{table*}

\section{Observations and data analysis}

Observations of two asteroids  Iris and Juno  together with observations 
of the sunlight at twilight have been collected at the VLT by means of the  
UVES spectrograph  between December 2006 and January 2007 as reported   in Table~\ref{obsdata}. 
UVES is a two-arm crossdispersed echelle spectrograph with the possibility to use 
dichroic beam splitters and to record most of the optical spectrum with  
one observation (Kaufer et al. 2004). The  observations have been taken with the 
dichroic  mode, the ESO DIC1,  allowing simultaneous observations of the blue and red arms.  
This dichroic  has a cross-over wavelength at 450 nm and the central wavelengths 
were set at 390 nm  for the blue and at 580 nm for the red arms respectively, 
allowing a full spectral range from 350 nm up to 680 nm. 
 
 We used a 0.5 arcsec slit  providing a resolving power of about 
$\frac{\lambda}{\delta\lambda} \approx 80,000$ 
which is the maximum resolution that can be reached with still adequate sampling of the PSF.  
It is relevant to note that the target is centered on one of the spectrograph 
slits, the red slit in our case,  while there is no way to check  the optical centering 
on the blue arm slit, directly.  The slits were  aligned with the parallactic angle  
in order not to miss  light due to atmospheric diffraction. The two UVES arms are equipped 
with CCD detectors, one single chip in the blue arm and a mosaic of two chips in the red arm. 
The blue CCD is a 2K$\times$4K, 15 $\mu$ pixel size thinned EEV CCD-44 while the  
red CCD mosaic is made of an EEV chip of the same type  and  a  MIT/LL CCID-20 chip 
for the redder part of the spectral range. Each arm has two crossdisperser 
gratings working in the first spectral order.
   
Asteroids are apparently fast moving objects and the geocentric radial velocity changes 
by about 1 \ms\  in about 4 minutes typically, thus limiting   
the maximum exposure time  also with the largest telescopes. At the epoch  
Iris and Juno  were of 8 and 10.6 mag, respectively, and we  kept the exposures at 
300 and 900 s  achieving a signal-to-noise  between 100--200.  

The observations were bracketed by ThAr standard calibration lamps.  
Calibration and  science observations were carried on in the  attached mode 
to avoid the automatic resetting of the spectrograph position, 
implemented by ESO on 26 Dec 2001,  to compensate  thermal drifts in the  
dispersion direction  between daytime  calibration frames and  science observations. 
The automatic resetting of the instrument  allows   calibration frames to be taken 
in daytime with an economy in terms of  observing time,   but it  makes an accurate  
calibration problematic.

The data reduction have been performed by means of the UVES Pipeline in the MIDAS echelle context. 
The wavelength calibration has been performed using the new atlas of 
ThAr spectrum by Lovis \& Pepe  (2007), which increases the  laboratory wavelength 
precision by means of HARPS observations, with the line selection suggested by 
Murphy et al. (2007) to avoid blends.  Mean
residuals of  $\leq$ 0.37 m$\AA$ for the blue arm, $\leq$ 0.46 m$\AA$ 
for the red low and $\leq$ 0.55 m$\AA$ for the red up  are generally obtained providing  
a velocity accuracy  at the central wavelengths of 
$\leq$ 25 \ms  in the red and of $\leq$ 30 \ms in the blue as shown in detail in Fig.~1.
We note that these residuals are about one order of magnitude  smaller than those derived by 
Chand et al. (2006). They can be further improved for limited portions of the spectrum 
where the reduction is optimized as achieved by Levshakov et al. (2007).  
As shown by de Cuyper \& Hensberge (1998) an accuracy of 10$^{-2}$ of the pixel, 
which in UVES is $\approx$15 \ms,  is attainable for non blended ThAr lines with 
more than 10$^3$ detected electrons in the central pixels.  
However, the accuracy of the ThAr lines themselves is of the order of 10 to 100 \ms\ 
and this is not directly reflected in the residuals of the wavelength calibration. 
The reduced spectra have been normalized manually  tracing  the continuum by means of 
the standard MIDAS routine.

For the reduction of the  twilight spectra we  skipped the automatic sky subtraction 
and extracted the spectra from the calibrated frames manually by using the standard 
MIDAS echelle commands, using a slit height of 8 pixels to minimize effects due to 
the small curvature of the slit projection on the detector.   
To check the curvature effects we  extracted  2 spectra from the same
image with an extraction slit of 2  pixels  and offset by 3 pixels above and below 
the central position of the order.  The position of  spectral lines on the two 
extracted spectra did not reveal notable shifts due to curvature effects.  

The angular sizes of the two asteroids in the epoch of observations were   
of 0.278 arcsec and 0.263 arcsec, and always smaller than the night seeing.  
They are effective  point sources and the light follows the same path through 
the atmosphere, telescope and spectrograph not differently  from a QSO or other 
point-like sources. Thus with  asteroid observations the radial velocity accuracy  
could  be monitored along the echelle orders for the whole frame in a much better way 
than with the calibration lamp since the asteroid lightpath takes into account 
the atmospheric variations and the centering of the object on the slit.   
In particular, in the case of UVES observations which make use of the dichroic we 
can monitor the response of the two separate arms.  In the twilight spectrum 
the diffuse day-light illuminates  the slit uniformly  so that a  comparison 
between the radial velocity of the asteroid and the day-light  probes slit 
illumination effects on radial velocities. 

UVES is also equipped with an iodine absorption cell which can 
be inserted in the beam to obtain a dense grid of iodine absorption 
lines superimposed on the target spectrum. The iodine cell currently mounted on UVES  
produces a rich absorption line spectrum in the range 
490-640 nm.  Butler et al. (2004) achieved an accuracy of 0.42 \ms\ 
for UVES with observations of $\alpha$~Cen~A by means of a  series of 3013 spectra of 1--3 s  
exposures, but after correcting for trends and jumps.  
However, iodine cell is not well suited for measuring accurate positions of 
QSO absorption lines which fall very far apart, and we are not aware of its use for this  purpose.

\begin{figure}
  \includegraphics[width=9cm, angle=0]{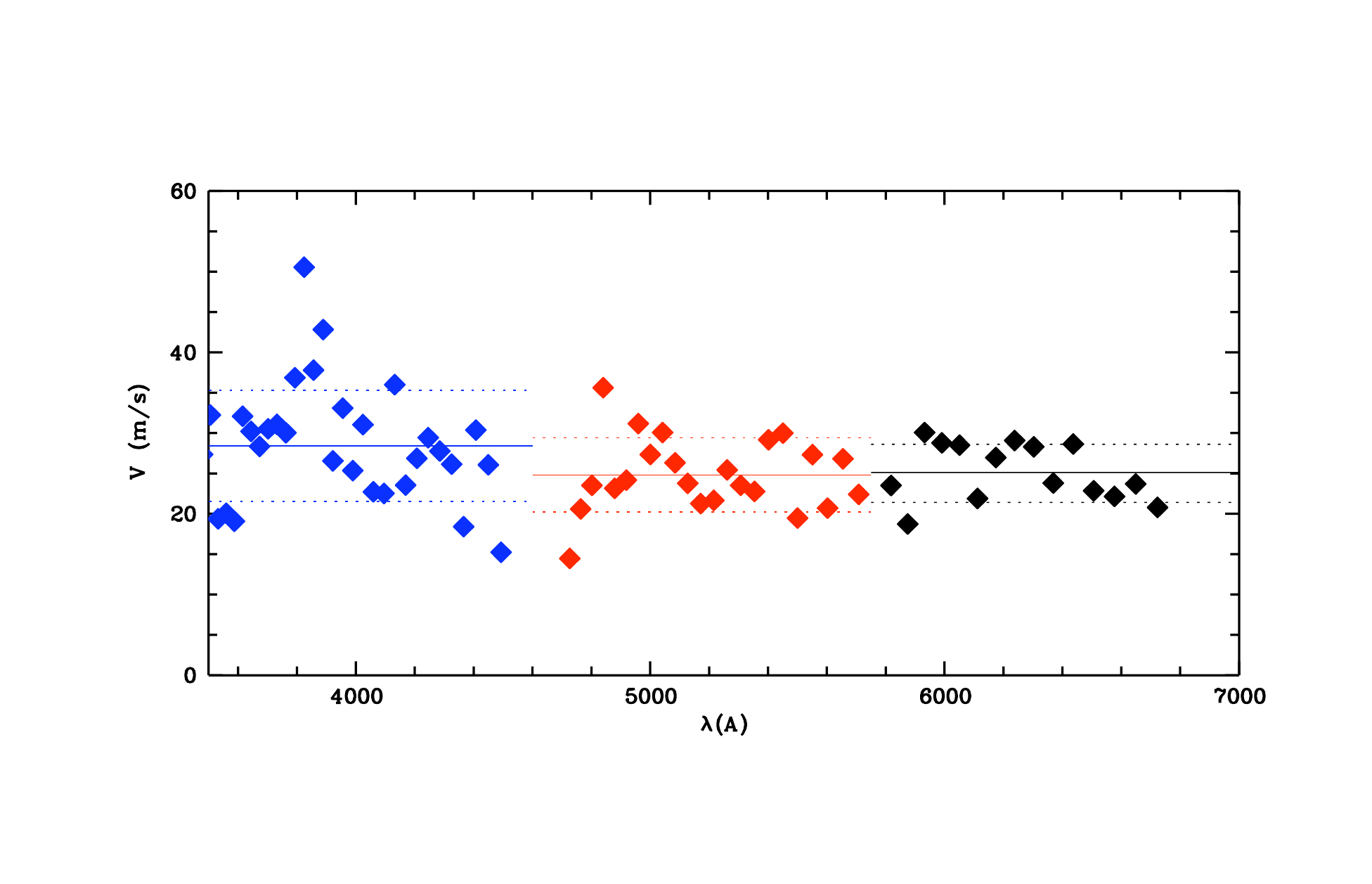}
\caption{Typical order residuals of the wavelength calibration, 
namely the difference between the measured and laboratory wavelength 
of the ThAr lines used in the calibration. The plotted ones are for  
Iris 22 Dec 2006.The three groups, with mean values and 1 $\sigma$ 
dispersion over plotted, refer to the 3 CCDs which have been reduced independently.}
              \label{fig1}%
   \end{figure}

\section{Asteroids as radial velocity standards}

Out of  111 stars observed in 20 years with the two CORAVEL spectrometers only 
a minor fraction shows variability of $\approx$200 \ms\  (Udry et al. 1999).  
Thus radial velocity standard stars provide a reference system of radial velocities 
with a precision of  several  hundred \ms. 
Among the celestial sources the asteroids are probably the  best  radial velocity 
standard sources and at least for two reasons. The former is that they   
reflect sunlight without any modification of the solar spectrum and the latter 
is that their velocity component with  respect to the observer can be predicted with 
very high accuracy reaching the \ms\ level (Zwitter et al. 2007).

The first condition  is strictly  valid only for relatively  large asteroids  
with  a  nearly spherical  shape which produce  a constant reflectance of the sunlight.  
The two selected asteroids Iris and Juno have radii of 99.9 and 117.0 km respectively and  
a spherical shape.  On the 18 Dec observation of Iris the illuminated fraction were 
of 97.07\%  and on the 24 Jan 2007 Juno had a 97.26 \% reflectance so that 
the reflected and the direct solar spectra are likely identical.    
Variation of reflectance with wavelength  or presence of regolith developed 
by meteoroid impact on the asteroid do not affect high resolution spectra.  
Also the asteroid rotation does not affect the solar spectrum and is  much smaller 
than the solar one. The rotational periods for Iris and Juno are of 7.14 and  7.21 hours    
respectively. Thus their rotational velocity would be of about 25 \ms,  
which is much  lower  then the solar one and it will not  cause  further significant broadening. 

The second reason is that  the component of their  motion relative to the 
observer on the earth  can be calculated with extreme accuracy.    
For asteroids with radar monitoring, the orbital computations 
can  take into account the interferences of  other bodies of the solar system 
including the major asteroids and reach precisions at the level of the \ms\ (Zwitter et al. 2007).

Table~1 reports  the motion components and the resulting expected radial velocity shifts 
$\Delta RV$. Ephemeris for our objects  has been computed by means of the JPL's Horizons  
system\footnote{Available at\ http://ssd.jpl.nasa.gov/horizons.cgi}   
which provides accurate ephemeris for the minor bodies of the solar system. 
The  sunlight  reflected by the asteroid is shifted by the heliocentric 
radial velocity of the asteroid with respect to the sun at the time t$_1$ 
when the photons left the asteroid and were shifted by the component of the earth 
rotation towards the asteroid at the time t$_2$, when the photons reach the earth.  
The latter is the projection along the 
line-of-sight of the asteroid  motion   with
respect to the observer  at the Paranal site  adjusted  for aberration,  
and  comprises both the radial velocity of the asteroid and the component 
due to the earth rotation towards the line of sight.
At Paranal the observed asteroid radial velocity is 
\begin{equation}
\Delta RV =(RV_{ast-par} + RV_{ast-\sun})
\label{E1}
\end{equation}

We also take as reference several twilight spectra which are listed in Table~2.
The radial velocity of the skylight reflected by the terrestrial atmosphere 
is also shifted by the heliocentric earth radial velocity. 
This component, $RV_{par-\sun}$,  is given in the last column of Table~2. 
The precise position of the scattering of  solar light by the atmosphere 
is not known  but it should be within 10 km from the ground.  
In the next sections we will show that the scattered light from the atmosphere 
probably keeps the transversal motions of the atmosphere and therefore it is  
not  possible to predict its velocity with the desired accuracy.

  \begin{figure}
 \includegraphics[width=10cm, angle=0]{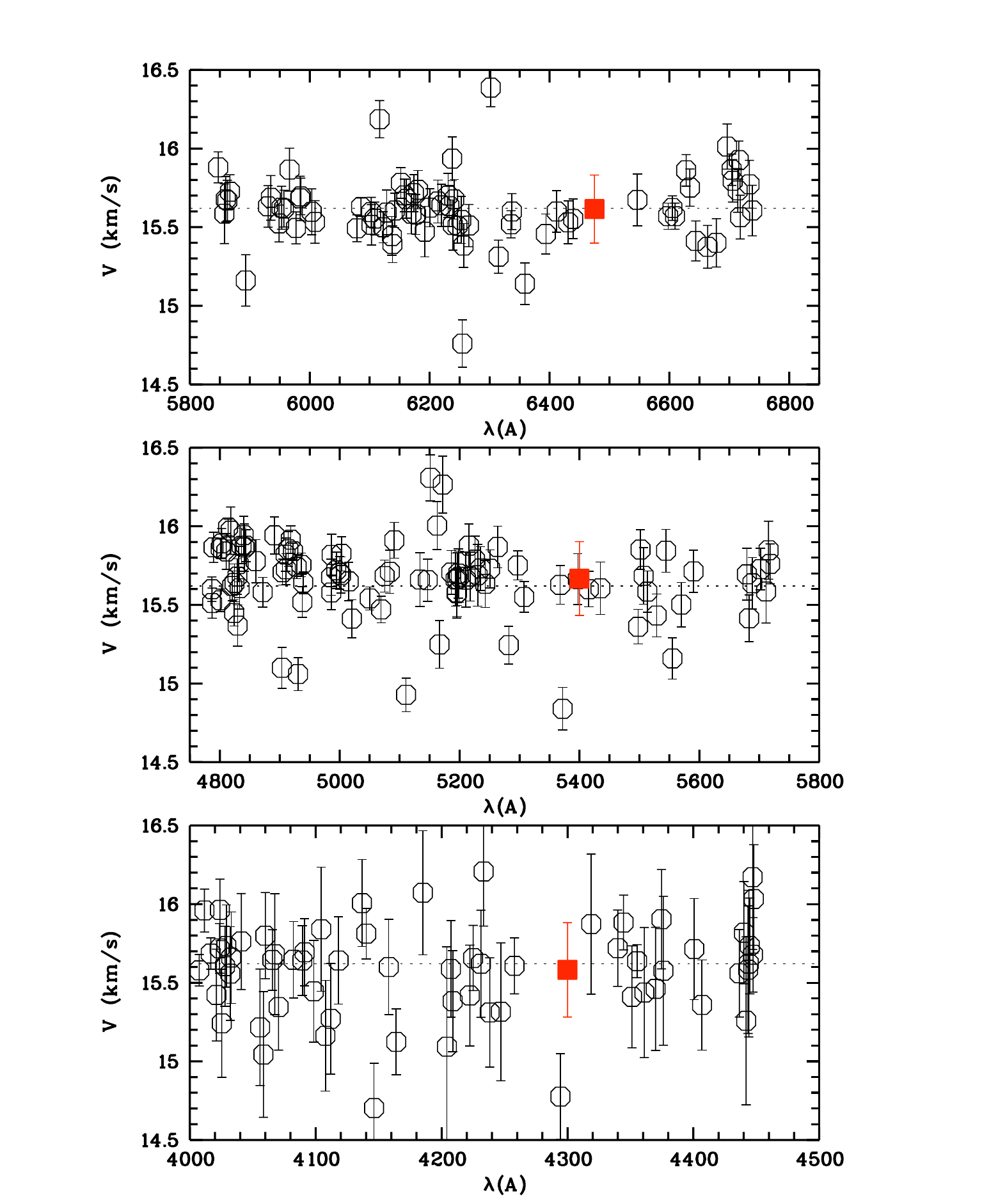}
\caption{Line shifts of Iris 23 Dec 2006    
with reference of the   Allende Prieto \&  Garcia Lopez  (1998a) 
solar line wavelengths, see text for details. 
The dotted line shows the expected velocity of the asteroid. 
The top panel refer to the red-up CCD, the middle panel 
to the red-low CCD, and the bottom panel the blue CCD. 
The  mean values and their dispersion are shown with the  squares   in the middle of each panel.}
     \label{Fig2}%
   \end{figure}


 \section{Asteroids with UVES}
 \subsection{Solar absolute reference}

The highest quality solar spectra in the optical domain are 
the FTS solar flux and disc-center atlas obtained at the McMath telescope at Kitt Peak  by 
Kurucz et al. (1984)  and Brault \& Neckel (1987). 
These atlases achieve a signal-to-noise ratio of about 2500 with  a resolving power of 400,000.
Allende Prieto \&  Garcia Lopez (1998a,b) used these atlases  to measure the central 
wavelength for a considerable number of lines.  Gravitational shifts and convective 
motions are responsible of line to line displacements which can be of several hundreds of \ms. 
These displacements vary with the solar cycle showing a modulation with a peak to peak 
variation of 30 \ms\ on the 11 years solar activity period with the positions more 
redshifted in correspondence of the maximum of activity (Deming \&  Plymate 1994).  However, 
McMillan et al. (1993) did  not reveal any drift  within 4 \ms\ in the solar line position  
from a long data series  spanning the period from 1987 to 1992.  
Allende Prieto \&  Garcia Lopez (1998a,b) line positions have a precision 
of the order of $\approx$50--150 \ms\  so that they provide  absolute reference at this level.  
The lines formed at the top of the photosphere show shifts close to the gravitational 
redshift of 636 \ms\ while the other lines show the effects of convective motions 
with variable blue shifts of several hundreds of \ms.    
Lines with  equivalent width stronger than 200 m\AA\,   
are rather insensitive to the convective shifts and have been  used to estimate 
the absolute zero of the scale.   
The value  at the plateau level is of $612\pm58$ \ms\ in the case of the solar atlas 
of Kurucz  et al. (1984) which shows the  results closest to the theoretical gravitational shift. 

We thus compare the measured line positions of the asteroid spectra with the 
solar line positions provided by  Allende Prieto \&  Garcia Lopez (1998a) for 
the solar atlas of Kurucz et al. (1984).  In fact the sun light reflected by the 
asteroids is a sort of  integrated solar flux as the Kurucz et al. (1984) atlases.  
Fig.~2 shows  the  \drv\ measures  for the Iris spectrum of  23 Dec 2006.  
The figure shows that there are no major wavelength calibration  inaccuracies 
at the level of 200 \ms\ which corresponds to about 0.1 of the pixels size.
The result shows  a mean value of 
\drv = $15.614\pm0.203$ \kms\  for the 75 lines measured in the red-up 
CCD,  a \drv = $15.668\pm0.234$  \kms\  for 96 lines in the red-low 
CCD, and of $15.582\pm0.300$ \kms\ for 63 lines in the blue CCD. 
Considering that the expected velocity 
is of \drv = 15.620 \kms, there is an excellent agreement with the  
red-up CCD and  a slight offset of about 50 \ms\  and 30 \ms\ with 
the red-low and blue CCD  respectively. Despite the scatter 
of $\approx$200--300 \ms,  
this analysis shows that there is no significant offset between the two arms 
of UVES  implying that there is no mis-centroiding  of the target    
on the two slits of  UVES arms.

\subsection{Asteroid versus asteroid}

To overcome the intrinsic uncertainties in the  positions of the solar lines 
we compare the solar spectrum from  an asteroid taken in two different epochs. 
In this way each line is compared with itself leaving only the instrumental and 
calibration imprinting on the change of the  asteroid  radial velocity between the two epochs.  

To measure accurately  the radial velocity difference between two lines   
we have  implemented  and adapted a procedure  from  Levshakov et al. (2006).  
The most probable  \drv\  between two lines is found by varying \drv\ 
by small incremental steps and estimating the  $\chi^2$ of the fit.
Fig.~3 shows a portion of the red-low frame of Juno 31 Jan and of the sky spectrum 
of the same day around  line 5217.09 $\AA$. The S/N are of 126 and 295 for Juno and 
twilight, respectively calculated from two nearby  continuum  windows bracketing the 
line position. The range used in the fitting is marked by thick curves on the upper 
panel of the figure and we consider only the central parts of the absorption 
lines to avoid the influence of the wings. 

From the fit of points in the vicinity of the global  $\chi^2_{\rm min}$ the procedure  
computes a parabola with the radial velocity difference as variable. The
1$\sigma$ uncertainty interval is then calculated from the parabola  when 
$\chi^2(\Delta v) -  \chi ^2_{\rm min} = 1$.
For this particular case we have obtained  \drv  = $-17.723\pm0.033$ \kms\ at 1$\sigma$.

The error is rather typical of our measurements and corresponds to about 0.02 of the pixel size.   
For instance, in computing the difference between the Iris spectra taken on  
18 and 22 Dec 2006,  the  203  lines measured  have a mean error of 
$39\pm3$  \ms,  and of  $37\pm2$ \ms, respectively.
The error is mainly  the photon noise error and  it depends from the signal 
to noise ratio of the two spectra. This error  sets  the precision of  
our analysis and the  level of instrumental effects which can be recovered.  
In principle with higher signal-to-noise spectra this level can be further improved.
Wavelength calibration errors are not expected to contribute very much 
to this error because even if the wavelength calibrations of two  spectra are 
performed independently, they likely make use of  the same Ar or Th lines  
in deriving the calibration coefficients. 
 
The results of the radial velocity difference between the Iris spectra taken  
on 18 with those of 22 and 23  Dec 2006 are shown in Figs.~4 and 5.  
The measures are performed on lines falling onto  7 orders for each CCD frame 
selected  to map the full spectral range.  On the right side of each panel the 
average value for each single order is reported with the sample standard deviation. 
As it can be seen from the top panel of the figure, the measurements do not  
show evidence for trends within an individual order, and the measures are normally 
distributed  around their  mean value. At the bottom of the figure the mean values 
for each order are plotted as a function of the order number. There is no evidence 
of any pattern of the measured radial velocity with  wavelength from  3500 $\AA$  up  
6750 $\AA$  with measurements involving 3 CCDs and two spectrograph arms.  
For the 18-22 comparison the mean of the 3 CCDs are   
\drv = $1.157\pm0.073$  \kms\  for the blue,  $1.140\pm0.048$ \kms\  
for the red-low  and  $1.134\pm0.048$ \kms for the red-up.  
The excess in the dispersion observed  within each order   
reflects  the combined contribution of wavelength calibration and data reduction 
errors with the statistical error. 
The mean of the mosaic of the two CCDs of the red arm   
is  only 19 \ms\ away from the value of the blue arm.  For the comparison between  
18 and 23 Dec observations we have a mean value for the blue CCD of $1.487\pm0.042$ \kms\ 
and for the mean of the two red CCDs a value of $1.464\pm0.036$ \kms, 
or 23 \ms\ away from the blue arm. 

Therefore, there is no evidence for a significant misalignment between the 
two arms of the UVES spectrograph. 
However, the expected radial velocity  difference between the two  epochs in which 
the asteroid was  observed is of  1.190 \kms\  and of 1. 447 \kms, respectively.   
Taking the mean value of the two arms  we  miss the expected velocity  by  
43 and  24  \ms\  in the two comparisons. This  implies  a sort of systematic error 
in one or in all the observations. The origin of this systematic error is likely 
to be ascribed to a  non-uniform  illumination of the slit.

  \begin{figure}
  \includegraphics[width=10cm, angle=0]{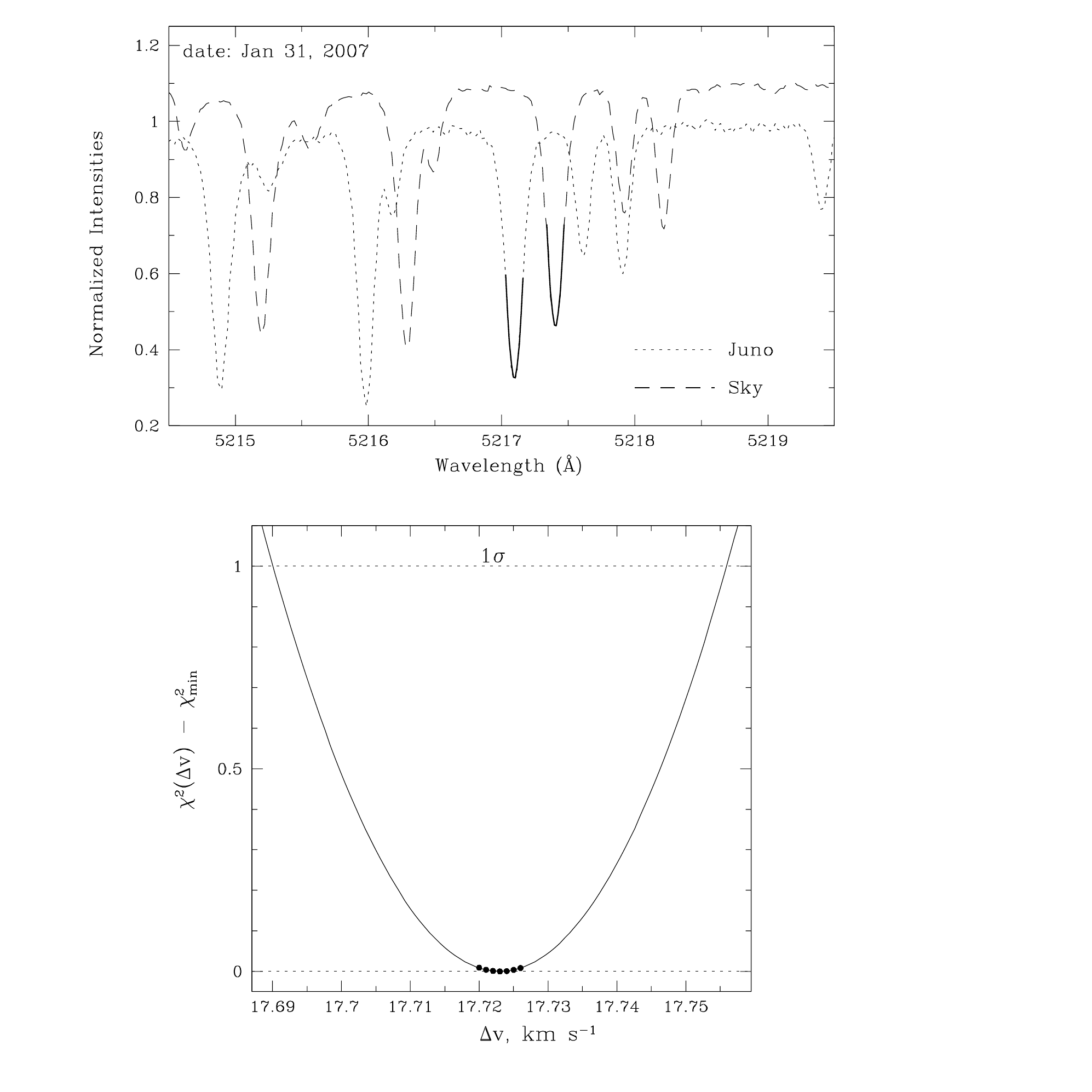}
\caption{Procedure for the determination of an accurate 
\drv  and its error. On the top panel the normalized sky 
spectrum is slightly shifted vertically for display purposes.
}
              \label{Fig3}%
   \end{figure}

Given that there is  no evidence for a systematic behavior within the orders  
in the rest of our observations we have performed a cross-correlation  to get  
order  shifts by means of the  IRAF-rvsao XCSAO routine.  For this kind of analysis 
particular care has been adopted in selecting   spectral regions without  
telluric lines which perturb the  cross correlation.   At the bottom of Figs.~4 and 5 the 
measures based on single lines, plotted in dots, with those performed by means of the XCSAO, 
plotted in diamonds, are showing that the two procedures are  providing consistent results.  
The results are reported in  Table~3 and summarized in Table~4. These measurements performed 
on the whole set of observations at our disposal  confirm  that there are not  notable 
patterns with wavelength, no offsets between the two UVES arms  and that  offsets  
with respect to the expected velocity in the range  10--50 \ms\ are common.

 \begin{figure*}
  \includegraphics[width=16cm, angle=0]{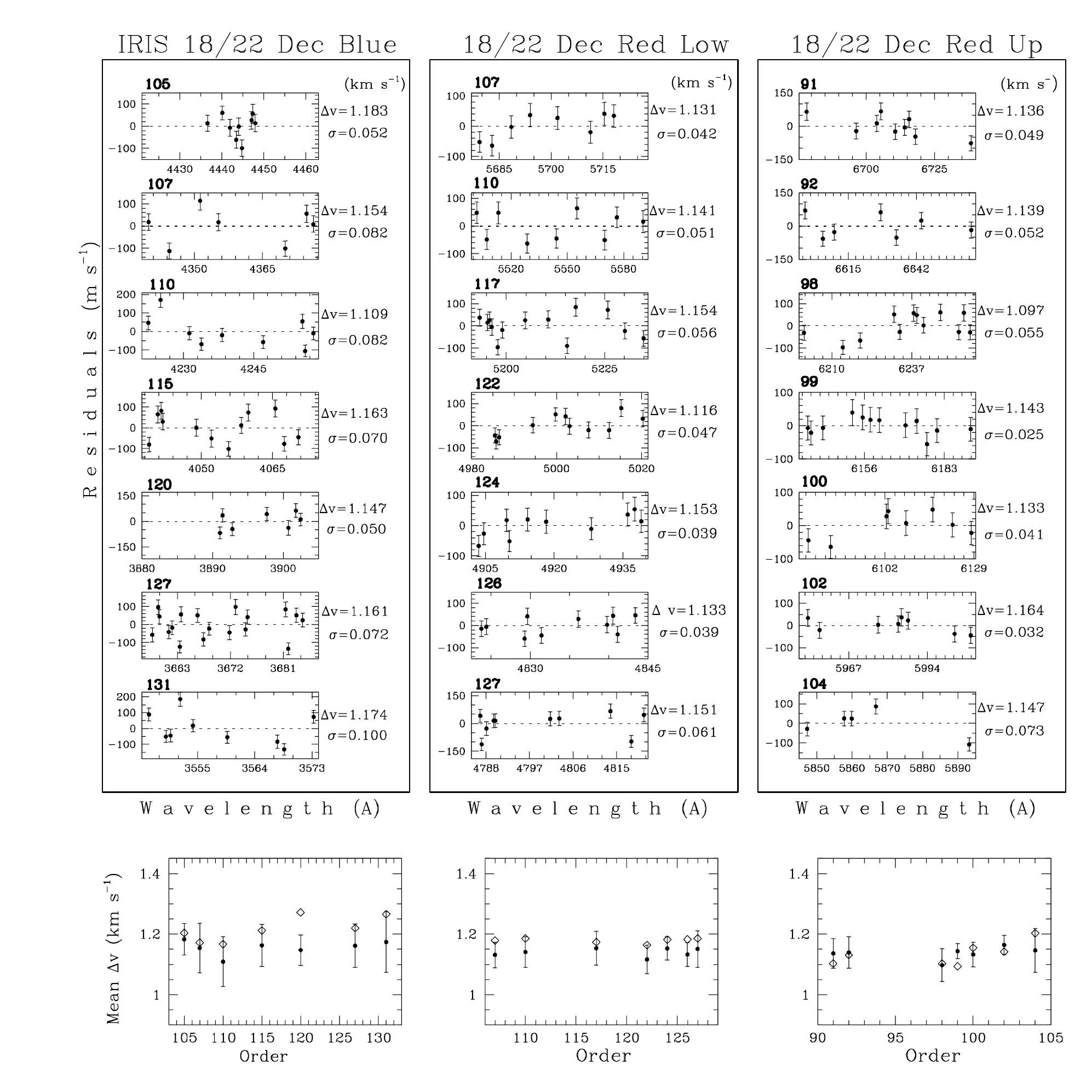}
   \caption{Radial velocity difference from Iris 18 and 22 Dec 2006. Residuals
     correspond to the difference Iris(22) -- Iris(18).
     The predicted $\Delta RV$ is 1.190 \kms. Individual echelle
     orders (numbered by bold) are shown in the upper panels. For each order the
     mean value $\Delta v$ and the sample standard deviation $\sigma$ are
     indicated. These values $\Delta v$ and  $\sigma$ are also shown
     by dots with error bars
     in the corresponding low panels where results obtained through the
     cross-correlation analysis (diamonds) are plotted for comparison.}
              \label{Fig4}%
    \end{figure*}

 \begin{figure*}
  \includegraphics[width=16cm, angle=0]{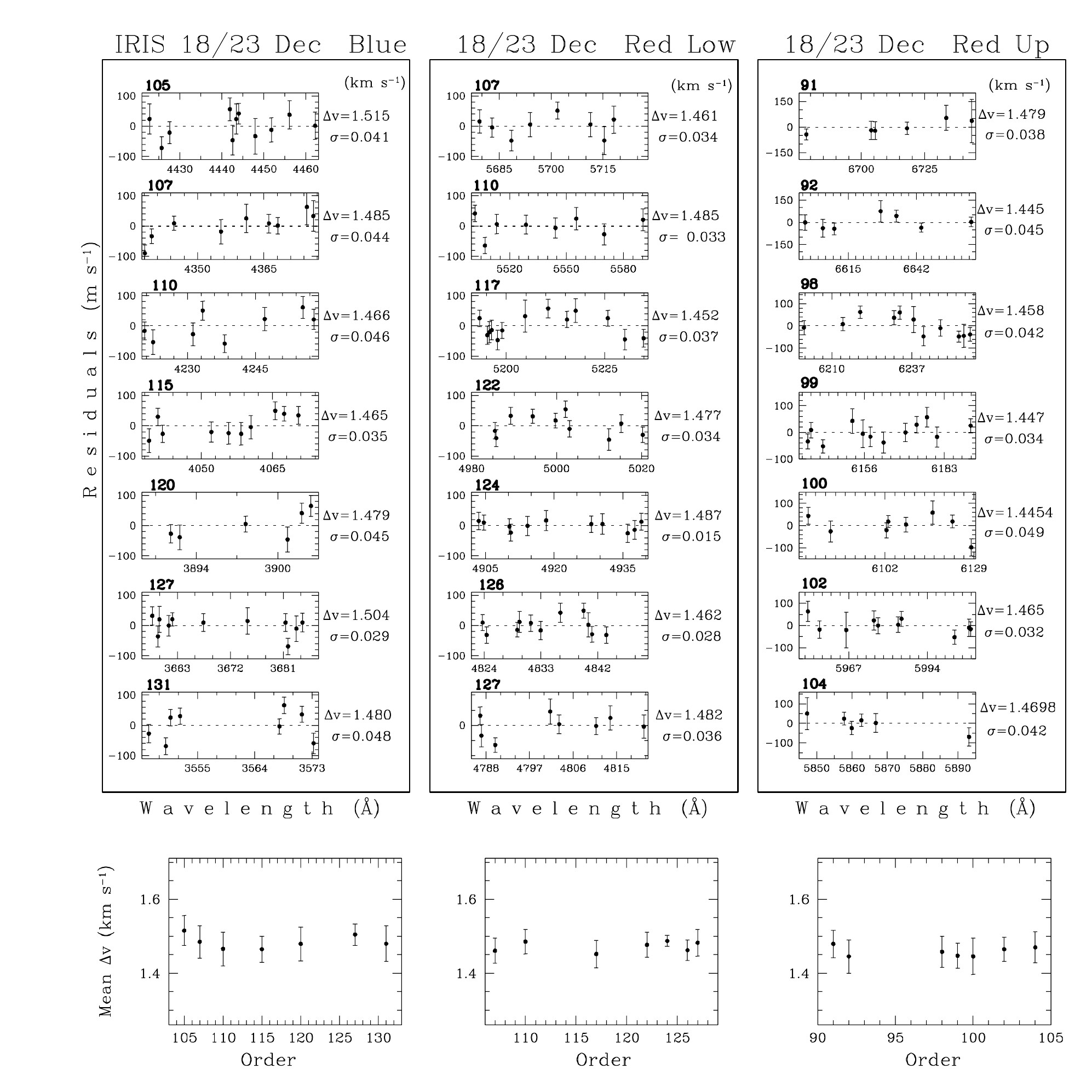}
   \caption{Same as Fig.~4 but for the comparison between Iris 18 and 23 Dec 2006.
        The predicted $\Delta RV$ is 1.447 km s$^{-1}$.
        Residuals correspond to the difference Iris(23) -- Iris(18). }
              \label{Fig5}%
    \end{figure*}

  \begin{figure}
  \includegraphics[width=12 cm, angle=0]{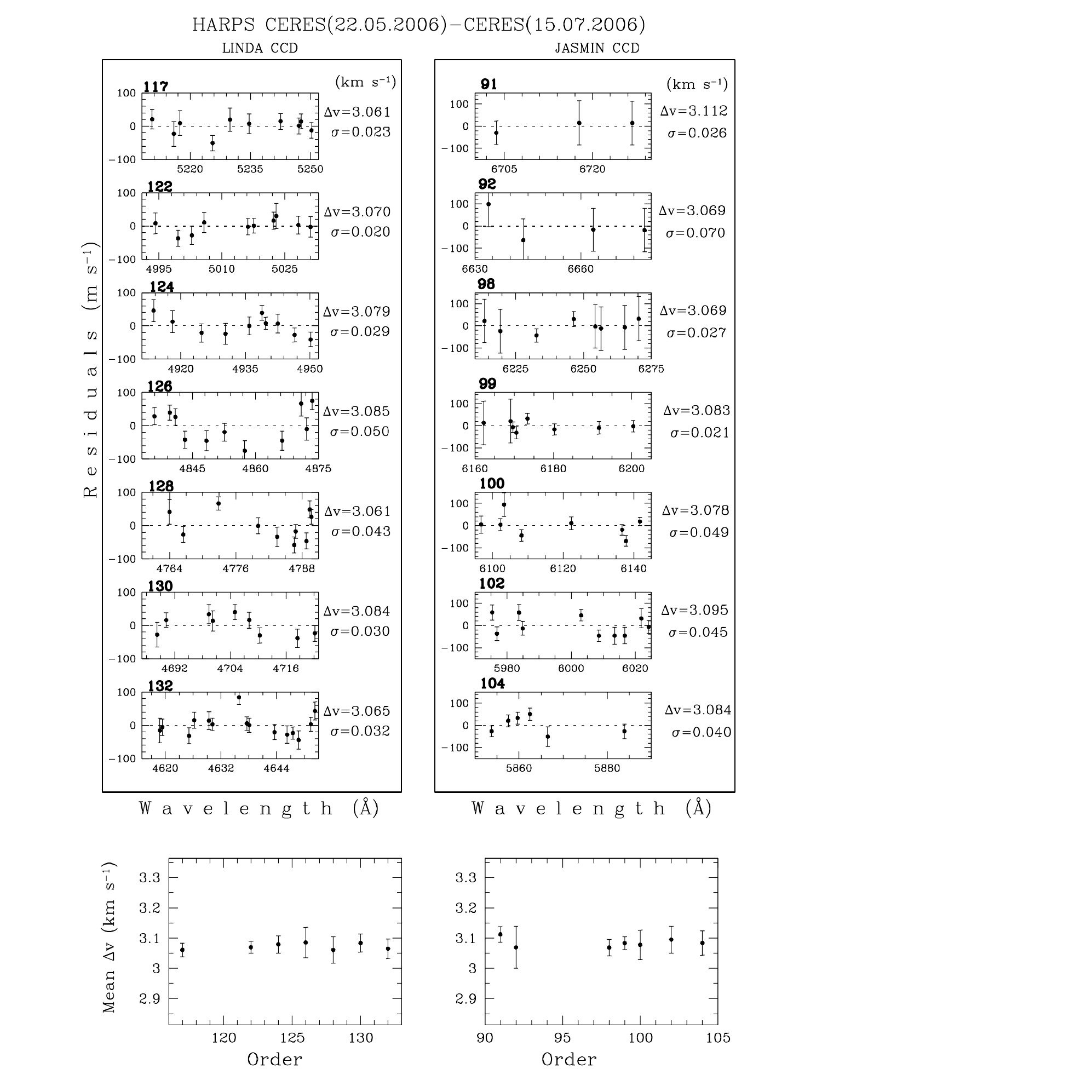}
   \caption{\drv  between Ceres  22 May   and 15 July 2006 . The expected \drv is  3.070 \kms}
              \label{Fig6}%
   \end{figure}

\section{Asteroids with HARPS }

To check the whole procedure by means of a different instrument  specifically 
designed for high precision radial velocity studies, we  retrieved  two reduced 
spectra of Ceres from  the public  HARPS archive and applied the same kind of 
measures performed with UVES. The High Accuracy Radial velocity Planet Searcher  
at the ESO La Silla 3.6m telescope is a spectrograph dedicated to the discovery 
of extrasolar planets through radial velocity oscillations. It is a fibre-fed high 
resolution echelle spectrograph
and  is contained in a vacuum vessel to avoid spectral drift due to temperature 
and air pressure variations.
There are two fibers, one  collects the star light, while the second is used 
to record simultaneously a ThAr reference spectrum. Both fibres are equipped 
with an image scrambler to provide a uniform spectrograph pupil illumination, 
independent of pointing decentering. In this way the instrument is able  
to obtain a  long term radial velocity accuracy of  the order of 1 \ms\ 
for the entire optical spectrum of a slow rotating G type star or cooler (Pepe et al. 2005).
HARPS has a resolving power of R~$\approx$120,000 and provides  a sampling 
of the slit of FWHM = 4.1 pixels of  15 $\mu$ size.  
Due to the relatively smaller size of the telescope  the exposures are  rather long,  
being namely of 1800 s and 900 s  (see Table~1).
In the course of the exposure the radial velocity of the asteroid changes 
by $\approx$50 and 25 \ms\ respectively. The expected velocities reported in Table~1  
refer to the mid exposure times.
  
In Fig.~6 the radial velocities measured between the observations  of Ceres taken on 22 May 2006 
and on 15 July 2006 are given. The accuracy of the measure of a pair  shift is now 
better than $\approx$20 \ms\ and the line to line variation of the positions is almost 
entirely due to errors in the wavelength calibration.
 The mean of the  blue CCD Linda  is \drv = $3.072\pm0.010$ \ms, 
and the mean of the red CCD Jasmin  is $3.080\pm0.010$ \ms.  
The predicted \drv\ shift is of 3.070 \kms\  and is found in excellent agreement 
with the  measured velocity within few  \ms .    
This is highly suggestive that the systematic offset observed in the UVES spectra 
is related to the slit acquisition mode which remains the most significant 
observational and technical difference between the two spectrographs.

\section{Twilight}

Sky observations differ from  point source observations mainly because the slit 
is uniformly illuminated.  Thus in principle a differential measure of a point-like 
source as  an asteroid with the same feature from the sky spectrum allows us to probe  
radial velocity drifts induced by a non uniform slit illumination. The results of the 
measures for the Juno observations of  31 Jan are shown in  Fig.~7.   
The  observations of Juno on  31  Jan when compared with the skylight  
on 31 Jan show that the \drv  inferred from the three CCDs  are  all consistent with each other. 
The blue CCD gives  a mean value of  
\drv = $-17.797\pm0.064$ \kms,   the red-low a \drv  = $-17.757\pm0.050$ \kms,  
and  the red-up a \drv = $-17.772\pm0.057$ \kms.  
The expected velocity is of \drv = $-17.914$ \kms, therefore we observed  an offset of
$\approx$150 \ms. 
This  offset is rather  high and about a factor three  higher than that observed 
in the series of asteroid-asteroid comparison.
To check the procedure we performed two separate tests. In one test we compare the accurate 
HARPS observations of Ceres with a sky spectrum taken with the same instrument, 
and in a second test we compare two twilight spectra taken with UVES in two different epochs.

Fig.~8 shows the  comparison between the spectrum of Ceres taken on  15 July  2006 
with a sky light taken with the same instrument on  22 Oct 2005. 
The mean \drv\  values in the two HARPS CCDs   are   \drv = 23.657  \kms\ 
for the blue CCD Linda, and 
\drv = 23.647 \kms\ for the red CCD Jasmin, while the predicted one is of  
\drv = 23.709 \kms\  computed for the  middle Ceres's exposure. 
This measure shows  an offset of $\approx$50 \ms\ between the Ceres and sky spectrum.  
This offset is not observed  when the Ceres of two epochs are compared with each other,  
as we  discussed in the previous section.
HARPS  is fed by fiber optics and therefore no difference is expected between 
the two kind of measures suggesting that  the sky  spectrum holds  a component 
of motion of  several  tens of \ms.  This implies that  the twilight  solar spectrum 
is not a good reference for the determination of the zero scale.

As a second test we  compared the sky spectra with each other.  
The difference between the sky spectra taken with HARPS on  22 May 2005   
and  14 July 2006  are shown in Fig.~9. The expected velocity difference for this pair is of 
$-371$ \ms\ while the mean value  is  $-275\pm39$ \ms. Thus, also in this case  
we fail to reproduce the expected velocity confirming that  the sky spectrum 
is sensitive to unpredictable motions likely due to currents in the upper  terrestrial atmosphere.  

We also emphasize that  close inspection of  UVES twilight and asteroid solar spectra 
show that they are not completely identical.  Small differences at the level of 
1-2\% are found between the twilight spectrum and the asteroid reflected solar spectrum  
consistently with what found by Zwitter et al. (2007). An example of the two spectra,  
with the skylight lines shallower, is shown in Fig.~10.
Similar differences have been found also by Gray et al. (2000)  and also depending on the  
angular separation from the Sun. According to Gray et al. (2000) the skylight variations can 
be explained as a combination of  Rayleigh-Brillouin scattering with a second term of aerosol.
The measure of the FWHM for a representative sample of lines from asteroid spectra and 
for  twilight are shown in Fig.~11. The FWHM of the twilight are significantly larger by 
about 5 m$\AA$ with comparison to the asteroid lines. The  broader  twilight lines 
suggest the presence of turbulence in the atmospheric motions which reflect the sun light.  
At twilight a transverse motion in the  atmosphere has a considerable component in 
the direction of the sun which is low above the horizon when the observations were made  
and produce a radial velocity drift when observed in the reflected spectrum. 
A detailed  investigation of these effects is beyond the scope of this paper, 
but the presence of this effects shows that the twilight spectrum is not a good  
zero reference point  at the level of  $\approx$100 \ms.

   \begin{figure*}
  \includegraphics[width=16cm,angle=0]{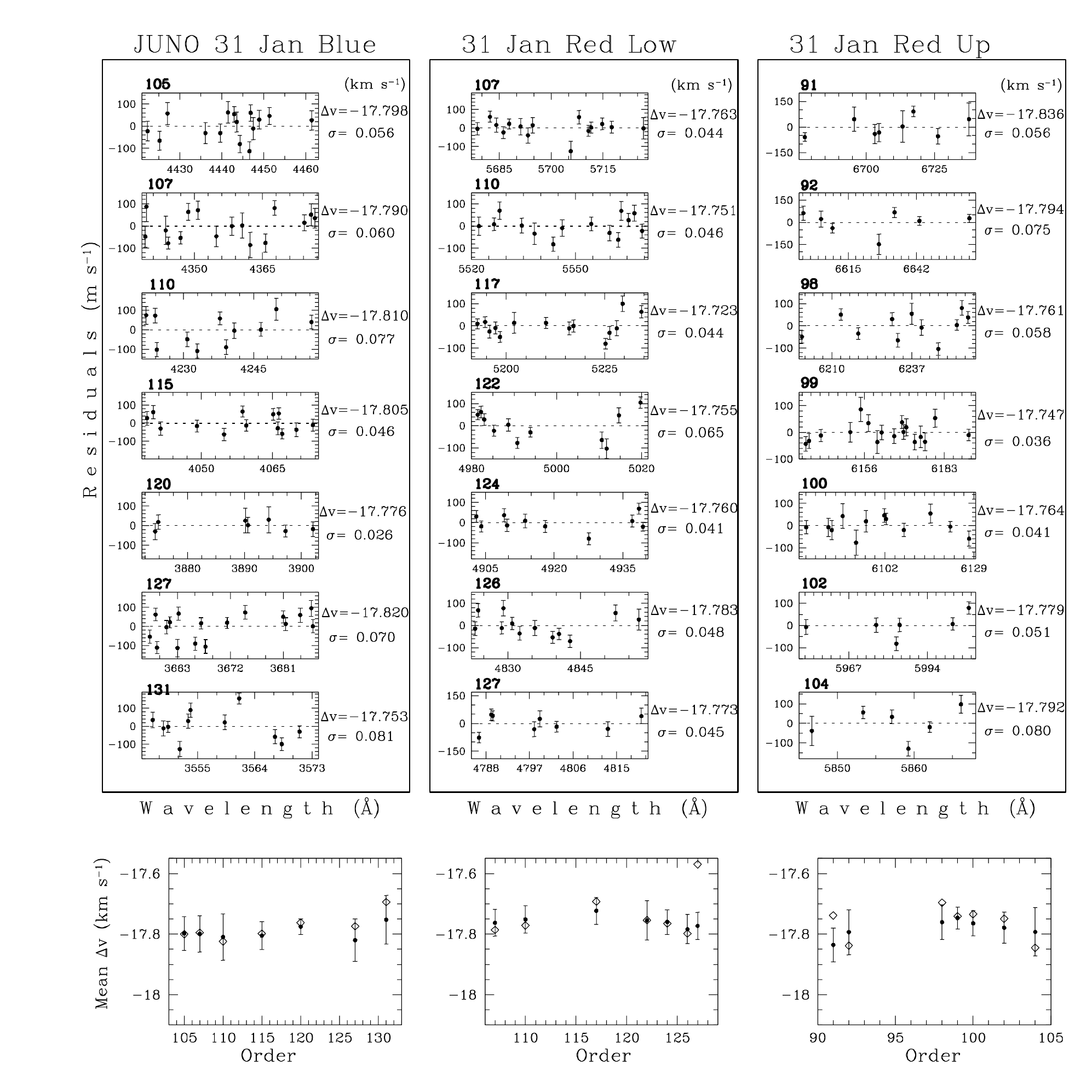}
\caption{Juno 31 January relatively to the twilight of the same date. Residuals correspond
to the difference Juno -- Sky. }
              \label{Fig7}%
    \end{figure*}

  \begin{figure}
  \includegraphics[width=12cm, angle=0]{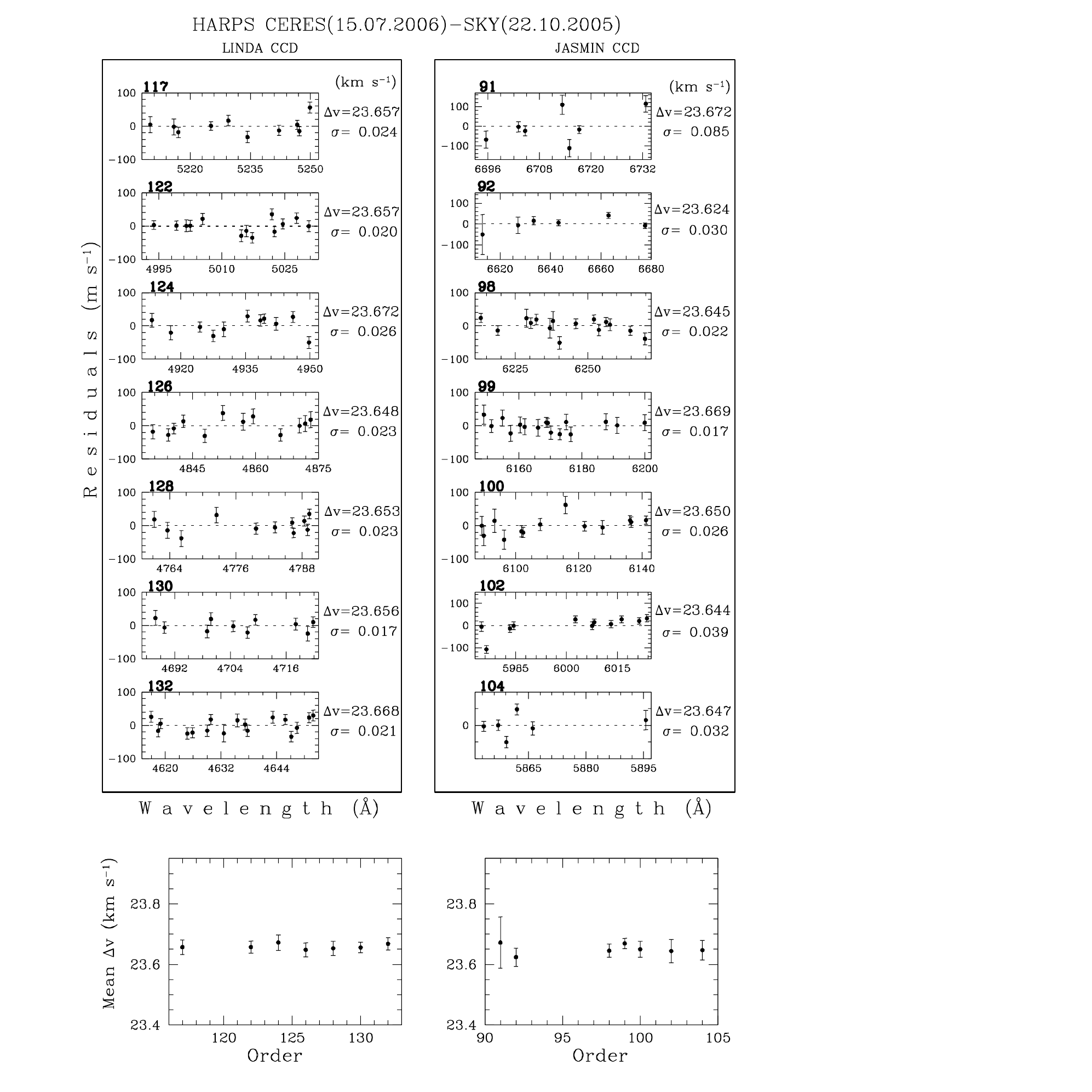}
\caption{\drv\ between Ceres  15 July 2006 and the Sky on 22 Oct 2005 . 
The predicted \drv\  is  23.709 \kms. }
              \label{Fig8}%
   \end{figure}

   \begin{figure}
  \includegraphics[width=6cm, angle=-90]{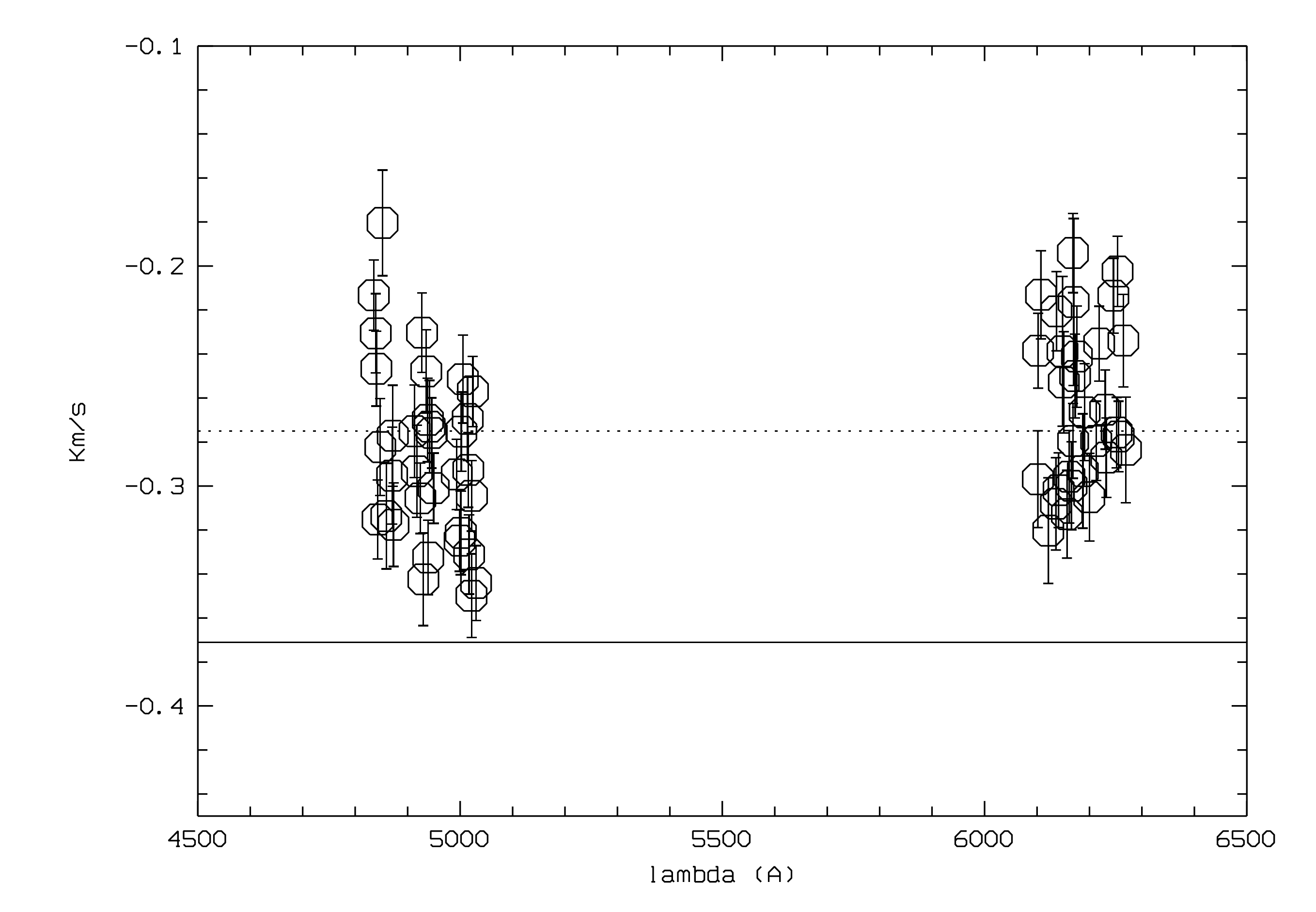}
\caption{\drv\   between HARPS sky of 22 May 2005  and  14 July 2006. 
The dotted line is the mean value while the red line shows the expected \drv\ at $-371$ \ms.}
              \label{Fig9}%
    \end{figure}
    
  \begin{figure}
  \includegraphics[width=6cm, angle=-90]{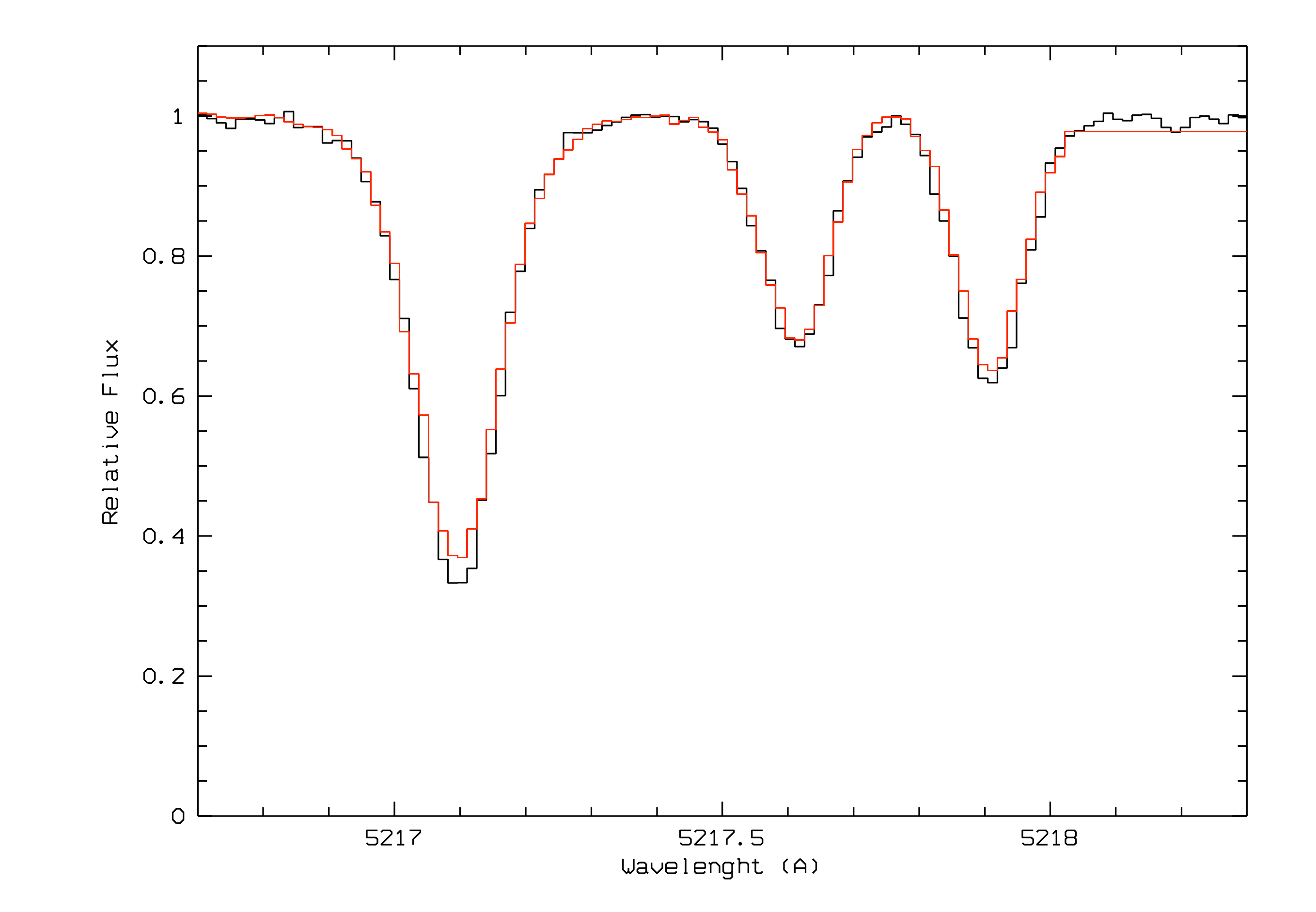}
\caption{Portion of the asteroid (thick line histogram)  
and twilight (thin line histogram) spectrum around the line 5217.3 \AA.}
              \label{fig10}%
   \end{figure}

   \begin{figure}
  \includegraphics[width=11cm, angle=0]{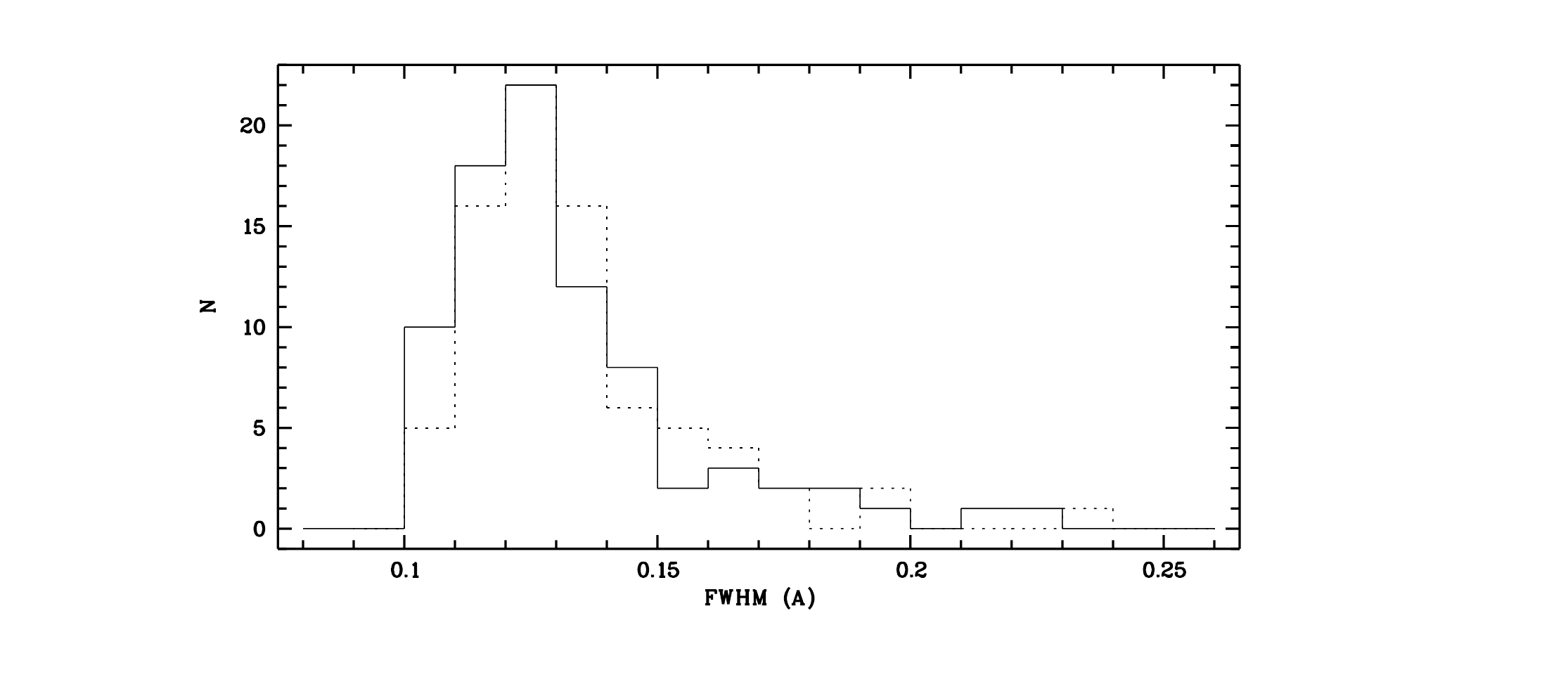}
   \caption{FWHM for asteroid Juno (solid  line) and twilight (dashed line).}
              \label{Fig11}%
    \end{figure}

 \begin{table*}
\caption{\footnotesize  Cross correlation analysis: differential radial velocity shifts
measured in \kms with respect of Iris 18 Dec 2006. The first column reports the echelle 
orders for the three chips blue, red-low and red-up. Xcsao $\sigma$  are also indicated.} 
\label{rvdataI}
\begin{tabular}{llclclclcl}
\hline\hline \\
Iris &     &  \multicolumn{2}{c}{18-22/12   } &  \multicolumn{2}{c}{18-23/12} & \multicolumn{2}{c}{18-24/12}  &  \multicolumn{2}{c}{18-25/12}   \\
B  & range $\AA$ & \kms &  $\sigma$ & \kms &  $\sigma$ & \kms  &  $\sigma$ & \kms  &  $\sigma$  \\ 
\hline                
105    & 4420.7 - 4463.0 &   -1.204     & 0.027  & -1.447  & 0.008  & -1.704  & 0.010  & -2.091  &  0.009 \\
107    & 4338.5 - 4379.2 &   -1.172     & 0.046  & -1.450  & 0.008  & -1.706  & 0.010  & -2.085  &  0.009 \\
110    & 4220.7 - 4259.2 &   -1.167     & 0.028  & -1.441  & 0.006  & -1.776  & 0.008  & -2.070  &  0.007 \\
115    & 4038.0 - 4073.3 &   -1.212     & 0.018  & -1.438  & 0.007  & -1.764  & 0.009  & -2.055  &  0.010 \\
120    & 3888.0 - 3902.8 &   -1.272     & 0.030  & -1.476  & 0.012  & -1.761  & 0.012  & -2.051  &  0.011 \\
127    & 3650.0 - 3686.9 &   -1.220     & 0.025  & -1.501  & 0.012  & -1.715  & 0.024  & -1.999  &  0.014 \\
131    & 3540.0 - 3573.9 &   -1.266     & 0.055  & -1.498  & 0.060  & -1.643  & 0.071  & -2.036  &  0.063 \\
\hline 			     		   		    
\\
RL \\			     	   		    
\hline			     		   		    
107    & 5675.2 - 5728.5 &   -1.179    & 0.029  & -1.462  & 0.007 & -1.705  & 0.011  & -2.095  & 0.008 \\
110    & 5521.1 - 5571.6 &   -1.186    & 0.025  & -1.452  & 0.009 & -1.774  & 0.013  & -2.097  & 0.009 \\
117    & 5192.2 - 5236.8 &   -1.174    & 0.021  & -1.403  & 0.007 & -1.763  & 0.011  & -2.071  & 0.012 \\
122    & 4981.0 - 5005.0 &   -1.157    & 0.008  & -1.465  & 0.006 & -1.717  & 0.008  & -2.103  & 0.007 \\
124    & 4900.3 - 4923.3 &   -1.167    & 0.009  & -1.468  & 0.007 & -1.749  & 0.009  & -2.102  & 0.009 \\
126    & 4822.8 - 4845.0 &   -1.183    & 0.041  & -1.461  & 0.008 & -1.726  & 0.011  & -2.084  & 0.009 \\
127    & 4810.0 - 4822.8 &   -1.190    & 0.027  & -1.470  & 0.013 & -1.752  & 0.043  & -2.117  & 0.018 \\
127    & 4786.4 - 4805.0 &   -1.183    & 0.024  & -1.436  & 0.012 & -1.740  & 0.020  & -2.135  & 0.026 \\
\hline
\\
RU \\			     	   		    
\hline			     		   		    
91     & 6667.5 - 6741.2 &   -1.103    & 0.047 & -1.442  & 0.024 & -1.701 & 0.042 & -2.145  & 0.038 \\
92     & 6595.5 - 6667.6 &   -1.131    & 0.026 & -1.461  & 0.031 & -1.773 & 0.032 & -2.181  & 0.030 \\
98     & 6193.7 - 6257.3 &   -1.103    & 0.012 & -1.463  & 0.011 & -1.777 & 0.014 & -2.120  & 0.012 \\
99     & 6131.5 - 6193.7 &   -1.094    & 0.011 & -1.456  & 0.009 & -1.774 & 0.010 & -2.098  & 0.009 \\
100    & 6077.0 - 6131.5 &   -1.155    & 0.015 & -1.431  & 0.011 & -1.689 & 0.016 & -2.105  & 0.018 \\
102    & 5995.0 - 6010.7 &   -1.142    & 0.017 & -1.452  & 0.022 & -1.741 & 0.027 & -2.144  & 0.027 \\
104    & 5841.0 - 5885.0 &   -1.203    & 0.036 & -1.487  & 0.021 & -1.659 & 0.019 & -2.065  & 0.020 \\
\hline
\hline \\
Juno &     &  \multicolumn{2}{c}{18-24/01} &  \multicolumn{2}{c}{18-25/01} &  \multicolumn{2}{c}{18-29/01} & \multicolumn{2}{c}{18-31/01}\\
B  & range $\AA$ & \kms &  $\sigma$ & \kms  &  $\sigma$ & \kms  &  $\sigma$ & \kms  &  $\sigma$ \\ 
\hline                
105    & 4420.7 - 4463.0 &  32.203    & 0.035 &  32.058    & 0.031  &  31.849    & 0.031  &  31.706  &  0.033\\
107    & 4338.5 - 4379.2 &  32.282    & 0.034 &  32.067    & 0.030  &  31.787    & 0.030  &  31.657  &  0.030\\
110    & 4220.7 - 4259.2 &  32.249    & 0.051 &  32.044    & 0.049  &  31.839    & 0.051  &  31.648  &  0.050\\
115    & 4038.0 - 4073.3 &  32.224    & 0.035 &  32.034    & 0.032  &  31.792    & 0.092  &  31.709  &  0.033\\
120    & 3888.0 - 3902.8 &  32.136    & 0.067 &  32.026    & 0.067  &  31.848    & 0.066  &  31.674  &  0.070\\
127    & 3650.0 - 3686.9 &  32.183    & 0.041 &  32.090    & 0.041  &  31.859    & 0.046  &  31.656  &  0.049\\
131    & 3540.0 - 3573.9 &  32.154    & 0.082 &  32.050    & 0.077  &  31.783    & 0.087  &  31.623  &  0.074\\
\hline	 
\\
RL \\	 
\hline	 
107    & 5675.2 - 5728.5 &  32.157   & 0.032  &  32.046    & 0.033  &  31.742    & 0.029  &  31.614  & 0.028\\
110    & 5521.1 - 5571.6 &  32.112   & 0.049  &  32.022    & 0.048  &  31.794    & 0.046  &  31.584  & 0.047\\
117    & 5192.2 - 5236.8 &  32.086   & 0.037  &  31.995    & 0.034  &  31.789    & 0.038  &  31.577  & 0.037\\
122    & 4980.3 - 5021.3 &  32.129   & 0.042  &  32.078    & 0.045  &  31.705    & 0.036  &  31.567  & 0.037\\
124    & 4900.3 - 4940.0 &  32.137   & 0.102  &  31.898    & 0.068  &  31.690    & 0.074  &  31.558  & 0.105\\
126    & 4822.8 - 4845.0 &  32.123   & 0.106  &  32.013    & 0.101  &  31.739    & 0.103  &  31.648  & 0.105\\
127    & 4810.0 - 4822.8 &  32.002   & 0.160  &  31.832    & 0.157  &  31.652    & 0.149  &  31.415  & 0.137\\
127    & 4785.0 - 4805.0 &  32.064   & 0.113  &  31.985    & 0.046  &  31.716    & 0.099  &  31.579  & 0.111\\
\hline	 
\\
RU \\	 
\hline	 
91     & 6667.5 - 6741.2 &  32.209   & 0.118  &  32.059    & 0.097  &  31.765    & 0.109  &  31.558  & 0.075\\
92     & 6595.5 - 6667.6 &  32.134   & 0.110  &  31.932    & 0.098  &  31.698    & 0.083  &  31.481  & 0.073\\
98     & 6193.7 - 6257.3 &  32.109   & 0.040  &  32.012    & 0.041  &  31.809    & 0.036  &  31.658  & 0.035\\
99     & 6131.5 - 6193.7 &  32.112   & 0.045  &  32.020    & 0.041  &  31.795    & 0.038  &  31.630  & 0.038\\
100    & 6077.0 - 6131.5 &  32.197   & 0.063  &  32.051    & 0.059  &  31.824    & 0.052  &  31.589  & 0.050\\
102    & 5995.0 - 6010.7 &  32.184   & 0.049  &  31.978    & 0.042  &  31.741    & 0.048  &  31.610  & 0.037\\
104    & 5841.0 - 5885.0 &  32.172   & 0.064  &  32.087    & 0.064  &  31.721    & 0.063  &  31.601  & 0.063\\
\hline
\hline
\end{tabular}
\end{table*}

\section{Implications for  \daa}

In the Many Multiplet method the  measurability of \daa from observations 
of absorption lines in QSO spectra is based on the fact that the
energy of each line transition shows a different sensitivity  on a change of 
$\alpha$ (Webb et al. 1999).  Thus, the value of \daa\  depends on the measure of 
the relative radial velocity shifts, \drv, between
lines with different sensitivity coefficients.  
The relation between the radial velocities and \daa\ is (Levshakov et al. 2006):
\begin{equation}
 (v_2 - v_1) = 
2\,c\,({\cal Q}_1 - {\cal Q}_2)  \frac{\Delta\alpha}{\alpha}\, ,
\label{E3}
\end{equation}
where  $\cal Q$  is the  sensitivity coefficient   ${\cal Q} = q/\omega_0$, 
with $\omega_0$  being the frequency and 
$q$ the theoretical  so-called q-factor. 
The  $q$-factors  have been computed for the most important  UV  resonance  transitions  
by Dzuba et al. (2002) and for \ion{Fe}{ii} were re-calculated by Porsev et al. (2007).   
 
The largest $\Delta {\cal Q}$ is presently provided  by  the $\ion{Fe}{ii}$ resonance lines. 
By comparing the \ion{Fe}{ii}$\lambda1608$
with and \ion{Fe}{ii} $\lambda2382$ or $\lambda2600$ lines 
($Q_{1608} = -0.0166$, $Q_{2382} = 0.0369$, and $Q_{2600} = 0.0367$ from Porsev et al. 2007) 
we obtain $|\Delta Q| \simeq 0.053$ which is almost 
two times larger than that obtained  from a  combination of other transitions.  
In this case a  shift of $\approx$30 \ms\ between the \ion{Fe}{ii} lines corresponds 
to a  \daa\  of $\approx$1 ppm.

Levshakov et al. (2007) analyzed  \ion{Fe}{ii} profiles 
associated with the \zabs = 1.84 Damped Ly$\alpha$  system  from UVES observations 
of  the quasar \object{Q 1101--264}. 
The data represent one of very few spectra of QSOs obtained with  spectral resolution 
FWHM  of  3.8 \kms\ and S/N~$> 100$. In this work a shift of the relative radial velocity  
between the $\lambda$1608 and  $\lambda\lambda$2382, 2600 lines of  
\drv = $-180\pm85$ \ms\ was obtained.  
With the updated sensitivity coefficients from Porsev et al. (2007)
this shift in the radial velocity between the \ion{Fe}{ii} lines corresponds to a  
\daa = $5.66\pm2.67$ ppm.  
                   
The \ion{Fe}{ii} lines fall at $\lambda \sim 4566$ \AA\
and $\lambda \sim 6765, 7384$ \AA, respectively, quite far apart  in the two different 
UVES arms so that  a  hidden systematic effect  would challenge the interpretation as 
due to variation of $\alpha$.   
Levshakov et al. (2007) measured the same velocity   between  the 
$\lambda 2382$ and $\lambda 2600$  which is what expected 
since the ${\cal Q}$ values for $\lambda 2382$
and $\lambda 2600$ are about equal. 
However, there is no direct way to check out systematic differences of the  
\ion{Fe}{ii} $\lambda1608$ with lines that fall  in the other arm of the spectrograph.
Different velocity offsets may occur in the blue and red frames causing an artificial 
Doppler shift between the \ion{Fe}{ii}$\lambda1608$ and $\lambda\lambda2382, 2600$ 
lines and mimicking  a change in \daa. The set of measures carried out here show that 
there are no \drv\ offsets between the two UVES arms  greater than 30 \ms. 
This  excludes  this kind of systematics as a possible origin of the signal detected 
by Levshakov et al. (2007). Therefore, either the detection is real or  it is induced 
by a  systematics of different kind.

\section{Conclusions}

Observations of asteroids have been conducted with the UVES spectrograph at the VLT  
to probe the radial velocity accuracy achievable with the spectrograph.  By means of HARPS  
observations we have shown that the asteroid observations are excellent radial velocity 
standards able to probe the  instrumental accuracy in any  particular position of the spectrum  
down to the limit  provided by the ThAr wavelength calibration, or 10 \ms.

By comparing the asteroid line positions with the absolute ones from solar positions 
which account for solar convective shifts we have shown that the UVES spectrograph 
is not affected by any systematics along the whole optical domain at the level where  
the solar line positions are  known, namely of few  hundreds of \ms.  We have further 
refined  the analysis by   comparing  of asteroid in different 
epochs. No major distortions in the wavelength are found, namely not higher  
than about 30 \ms, where this limit is set by the photon noise of our observations. 
We do indeed reveal zero offsets in the range  0  up to  $\approx$50 \ms. 
With reference to similar observations performed with HARPS we suggest that this 
is likely due to a non uniform slit illumination. 
Attempts to use the twilight spectra  to quantify the drifts induced by non-uniform 
illumination shows instead that twilight spectrum contains additional turbulence and motions, 
and therefore cannot be used as a reliable zero reference point.  

The recorded spectrum does not show evidence of stretching of the wavelength scale 
or other instrumental effects in excess of the uncertainties induced by the wavelength 
calibration accuracy. In particular, the two UVES  arms which are fed by two independent 
slits do not show signature for radial velocity offsets within the present accuracy of  30 \ms.

This result  has important  implications on the search for \daa\  currently performed 
with UVES which relies  on relative shifts of absorbing  lines  falling on  rather 
distant spectral regions and sometimes belonging to different arms of the spectrograph.  
For instance,  Levshakov et al. (2007) measure of  a \drv\ difference of 
$-180\pm85$ \ms\   between \ion{Fe}{ii} transitions falling in the two different arms of 
UVES  which provides  evidence for a
variation in the fine structure constant 
\daa = $5.66\pm2.67$ ppm. 
The present analysis shows that  the line shift is unlikely  produced by a misalignment 
of the the two slits at the entrance of the two UVES arms.

The proposed technique has a general validity and can be applied to any spectrograph 
to perform a real time quality control  of the spectrograph performance during night 
time while the observations are  carried on.

\begin{table*}
\caption{\label{rvdataI} \footnotesize Summary of mean radial velocity shifts
measured in \kms with respect of Iris  18 Dec 2006.} 
\begin{tabular}{llclclclclc}
\hline\hline \\
Iris \\
\hline
Date     & $\Delta RV$ & Blue &  $\sigma$ & Red-low &  $\sigma$ & Red-up &  $\sigma$ & Red  & Tot & $\sigma$ \\ 
18-22/12 & -1.196   &  -1.216  &  0.041      &  -1.179  &  0.011      &  -1.133  & 0.038      & -1.156    & -1.176    & 0.046 \\
18-23/12 & -1.481   &  -1.464  &  0.027      &  -1.474  &  0.063      &  -1.456  & 0.018      & -1.465    & -1.463    & 0.039 \\
18-24/12 & -1.765   &  -1.724  &  0.047      &  -1.750  &  0.039      &  -1.731  & 0.048      & -1.740    & -1.735    & 0.042 \\
18-25/12 & -2.034   &  -2.055  &  0.031      &  -2.092  &  0.015      &  -2.122  & 0.038      & -2.107    & -2.092    & 0.042 \\
\hline\hline \\
Juno \\
\hline 
Date     & $\Delta RV$ & Blue &  $\sigma$ & Red-low &  $\sigma$ & Red-up &  $\sigma$ & Red & Tot &  $\sigma$ \\ 
18-24/01 & 32.083   &  32.204  &  0.051      &  32.101  &  0.050      & 32.160   &  0.041     & 32.130    & 32.152    & 0.063 \\
18-25/01 & 32.054   &  32.053  &  0.021      &  31.983  &  0.087      & 32.020   &  0.052     & 32.002    & 32.017    & 0.063 \\
18-29/01 & 31.793   &  31.822  &  0.033      &  31.730  &  0.052      & 31.765   &  0.047     & 31.747    & 31.770    & 0.058 \\
18-31/01 & 31.696   &  31.668  &  0.031      &  31.568  &  0.068      & 31.590   &  0.057     & 31.579    & 31.606    & 0.068 \\
\hline\hline \\
\end{tabular}
\end{table*}

\begin{acknowledgements}
The asteroids observations were obtained in service mode in UVES calibration time. We 
are grateful to Cedric Ledoux  and to all UVES operation astronomers for the careful job which 
has made possible these measurements. We thank also Fiorella Castelli and  
Cristophe Lovis for many useful discussions. 
Part of this work was supported by  PRIN-INAF 2006.  
S.A.L. gratefully acknowledges the hospitality of ESO (Garching) and
Osservatorio Astronomico di Trieste. 
This research has been supported by
the RFBR grant No.~06-02-16489,
by the Federal Agency for Science and Innovations
grant NSh~9879.2006.2,
and by the DFG project RE 353/48-1.
\end{acknowledgements}

\end{document}